# Reply to Comment on "Negative refractive index in artificial metamaterials" (preprint arXiv.org:physics/0609234)

A. N. Grigorenko,

*Department of Physics and Astronomy, University of Manchester, Manchester, M13 9PL,*

*UK*

We discuss the claims of the comment at arXiv.org:physics/0609234. We show that A.V. Kildishev *et al.* misread our method of extracting of optical constants of nanostructured films. The theoretical calculations performed in the comment appear to be in a direct contradiction with an experiment. We demonstrate that the theoretical calculations suggested being free from ambiguities (V.M. Shalaev *et al.*, Opt. Lett. 30, 3345 (2005)) require an additional experimental verification, which can be performed by observing physical effects of negative refraction.

We thank Shalaev's group for the time devoted to our work [1,2] where we have demonstrated metamaterials with permeability $\mu'$=0.5 for double-pillar arrays without glycerol [1] and $\mu'$=-0.3 and $n'$=-0.7 for a sample covered with optically thin layer of glycerol [1,2]. We have also stated [1] that we did not observe negative refraction because of large dissipation in our samples. By negative refraction we understand a physical effect of light refraction at a negative Snellius angle (contrary to a negative value of $n'$ calculated with some theory for some theoretical medium.) In negative refraction experiments we attempted to measure a negative displacement of an oblique beam experiencing multiple reflections inside the sample without success due to large dissipation.

In the comment [3] and an unrefereed preprint [4] Shalaev's group disputes our results. It is not the first time V. M. Shalaev makes an attempt to undermine our priority in this rapidly developing area of nanooptics. At first, Shalaev's group tried to claim their priority in developing optomagnetic metamaterials for visible optics on the basis of an unrefereed preprint [5]. Now they dispute our results on the basis of their theoretical analysis of our data. As far as we understood, there are three main criticisms of our work.

1. Use of the Drude model.
We agree that the Drude model is not the best approximation for describing the behaviour of electrons in nanometre sized metal particles. We doubt, however, that the matching Debye model suggested in [4] is much better for modelling an electromagnetic response of a nanoparticle *of a general shape placed on any substrate and in any environment*.

This is a *theoretical* suggestion which requires *experimental* verification. At the same time, one can easily check that, for the symmetric and antisymmetric shape resonances of a pair of nanoobjects [6,7], the choice of the spectrum is of secondary importance and the resonance positions can be adjusted by the parameters of the Drude model. In addition, the choice of the electron model does not significantly change the calculated magnetic moment (provided the model adequately describes the plasmonic resonances). Therefore, a concern about the Drude model application is irrelevant.

2. Method of optical constant determination.

The comment [4] introduces a Shalaev's interpretation of our method (SIM) of extracting optical constants of nanostructured media. SIM is based on measurements of reflection spectra at only one (normal) angle of incidence and the comment [4] prove it to be wrong. We will not defend SIM (although we do not entirely agree with all claims [4]) simply because it was not the method that has been used in our analysis.

By an effective optical film for a nanostructured sample we understand a film of some thickness, permittivity and permeability which reflection and transmission coefficients give the best approximation to the reflection and transmission coefficients measured for a sample illuminated by light of an arbitrary angle of incidence. Therefore, optical constants of our samples should be determined and have been determined by a *combined* analysis of the reflection spectra and ellipsometric data, see the supplementary material of Ref. [1]. For this reason, the thickness of an effective optical film in our calculations is a parameter that should be determined from experiments. In the comment [4], on the other hand, the thickness of an effective optical film is predetermined and is given by the size of nanoparticles. This is obviously wrong. Indeed, two identical arrays of nanoparticles of different (small) sizes but of the same electric and magnetic susceptibilities would produce the same reflection and transmission coefficients illuminated by a light (of an arbitrary angle of incidence). As a result, these two arrays should have the same effective optical thickness despite having different sizes of nanomolecules, which contradicts to the approach [4].

In addition to the combined analysis of the reflection for p- and s-polarizations under oblique and normal angles, we have checked the extracted optical constants of our samples using interferometry (see below), which allows one to directly observe the phase of the transmitted wave and an optical path length in a sample.

3. Theoretical calculations [4] for our structures.

V. M. Shalaev and co-authors model our reflection spectra using their theory and find disagreement with our results. As we mentioned above, theoretical optical constants obtained from normal reflection spectra are obsolete unless they are checked with the reflection under some angle (or phase measurements), see the point 2. (SIM gives a large error for many experimental geometries). As we understand, the approach [4] could provide an effective optical film capable of modelling the response of our structures for a normal reflection (transmission) but most probably will fail for reflection (transmission) under an oblique angle. Unfortunately, it is impossible to judge on agreement of the model [4] with our ellipsometry data since the Brewster angle is not calculated nor the reflection near Brewster angle is given in [4].

However, it appears that the theory [4] contradicts to our interferometric measurements. It is very difficult to present the interferometric measurements on samples covered with optically thin layer of glycerol due to residual of glycerol near the structures, which affects the interferometric pattern. (We leave a discussion of such samples and their optical thickness to a later publication). Instead, we demonstrate interferometry of samples with higher density of nanopillars (the lattice constant of $a$=320nm). Reflection spectra of these samples mimic the spectra of the samples covered with glycerol. The parameters of nanomolecules for these samples are analogous to those used in [1]. Figure 1(a) depicts the standard interferometric scheme that we applied to measure an optical path length in the sample. This scheme utilises the standard Mach-Zander interferometer, Zoom 160 optical system and high resolution CCD. The interferometer is set in the fringe mode and the sample is placed in one arm of the interferometer. If an optical medium is present on a substrate, the optical path due to this medium contributes to the beam phase and the fringes bend. For glycerol, which refractive index is larger than that of the air, the phase difference lags and the fringes bend to the right. Figure 1(b) shows the phase lag of light passing through the glycerol droplet placed on a glass substrate. In case of the double-dot structures observed in red light ($\lambda$=632.8nm), Fig. 1(c), the refractive index of the sample is close to the refractive index of the air and the fringes do not bend at all. For this reason, Fig. 1(c) shows zero phase difference. In case of the double-dot structure observed in green light ($\lambda$=543.5nm), Fig. 1(d), the interferometric fringes bend in the opposite direction (to the left), which corresponds to a lead of the phase of light passing through the medium in relation with the air (the sample edges are shown by a grey box). This experimental result directly contradicts to Shalaev's group theory [4], where they obtained $n'>1$ for our structures at green light. The phase lead of Fig. 1(d) is large and corresponds to a negative optical path length for the light travelling through our medium (in our understanding of an effective optical film). Fig. 3(e) gives a zoomed version of the phase lead observed in green light. Fig. 1(f) shows the phase shift for our best sample (with larger nanopillars connected by a metallic bridge) at $\lambda=543.5$nm. Here the phase lead is about $\pi$ which results in a negative optical path length (in any model!) and corresponds to $n'<-1$ at the green wavelength. Measurements of a light phase for small objects require a lot of attention and we hope to present detailed analysis of these experiments soon.

To the best of our knowledge, this is the first experimental demonstration of a negative optical path length in visible light optomagnetic metamaterials (impossible if theory [4] is right). These interferometry experiments (and ellipsometry data) make us to believe that calculations [4] cannot be applied to our experiments. It is difficult to judge on reasons for this discrepancy. It could happen that [4] use an erroneous distribution of glycerol (glycerol should be placed just in between nanopillars [8]), or an influence of glycerol on surface states of electrons in a particle has been miscalculated. It is possible that the Debye model does not work for the double dots with glycerol in between them. It has to be said that [4] do not provide any details of calculations for the samples with optically thin layer of glycerol where we found negative $n'$.

We understand limitations and difficulties of theoretical analysis of such a complex system and we tend to believe experiments rather than elaborate theory. For this reason, when demonstrating non-trivial magnetic permeability [1], we have presented an *experimental* observation of impedance matching for non-zero magnetic permeability. It is suggested in [4] that the work on infrared structures [9] is based on "unambiguous" theory and therefore could be regarded as error-free. We do not share this view. We feel that the work [9] suffers from experimental and theoretical ambiguities and therefore requires an additional experimental verification.

1. Experiments.

The Shalaev's group claims that they measure absolute phases to determine complex transmission (and reflection) coefficients. It is well known, however, that an absolute phase is notoriously difficult to measure (especially for transmission) and one can get ambiguous results using such approach. Later in Ref. [9], V. M. Shalaev *et al.* state that they do not measure an absolute phase, instead they measure a relative phase of the transmitted signal. However, the phase of an optical signal coming from a small sample has a tendency to average with the phase from the other parts of a Gaussian beam, which gives an ambiguous phase result. Adequate determination of the relative phase requires some lenses, which are absent in the interferometer set-up shown in the preprint [10]. It also appears that Shalaev's group uses focused beams to measure an optical phase of a beam transmitted through the sample [10] which can itself contribute to a large phase error.

2. Theory.

We leave aside the question of determining 5 parameters of an effective optical film (Re($\varepsilon$), Im($\varepsilon$), Re($\mu$), Im($\mu$) and thickness) from just 4 measured experimental data (amplitudes and phases of the transmitted and reflected waves). We believe that such approach should be checked by confirming extracted optical constants using light of an oblique angle of incidence. There are 3 main difficulties with the theory [9].

i) Due to nature of the lithographic process, the infrared structure studied in [9] has the top stick that is notably smaller than the bottom one. As a result, the antisymmetric mode [6,7] in geometry [9] has a large contribution from an uncompensated dipole moment, which strongly affects the magnetic response. This is not discussed in [9].

ii) The infrared structures [9] do not posses a nice property of a local excitation of the magnetic antisymmetric mode by an electric field (in contrast to single-spit resonators [11] or our structures [1]) because of different geometry of the nanomolecules (the antisymmetric mode in [9] is a dark mode with respect to electric field). For this reason, it appears that the infrared structures [9] generate a non-local magnetic response, which does not lead to a standard permeability. Again, this is not discussed in [9] or anywhere.

iii) Since the magnetic response of the infrared structures is non-local, it appears that one needs to take into account the non-local contribution of a quadrupole electric mode which happens to have the resonance at the same frequency as the magnetic dipole mode and which gives a contribution to a non-local electric permittivity. We did not find a discussion of the quadrupole electric contribution in the works of Shalaev's group either.

These suggest that the theoretical result concerning negative *n′* [9] should be regarded with some amount of sound scepticism. We believe that any theoretical negative *n′* should be backed up with an experimental observation of negative refraction by which we mean a negative Snellius angle, a negative optical path length or a negative displacement of the beam passing through a negative index medium under some angle. The visible and near infrared spectrums have a nice property that the phase of light can be visualised. It would be of great benefit to metamaterial community if Shalaev's group could back up their own theoretical calculations of negative index of refraction [9] for the infrared system by *an experimental observation* of *any physical effect* connected to negative refraction (or at least non-trivial permeability!), e.g., a negative optical path length. Till then, we consider the claims about negative index of refraction made in [9] (and other works) as useful theoretical speculations.

To conclude, the comment [4] misinterpret our methods while the theoretical calculations [4] appear to contradict our experiments. We suggest that the calculations for the infrared structure [9] should be supported by an experimental observation of negative refraction. On behalf of all co-authors of Ref. [1], A. N. Grigorenko. e-mail address: sasha@man.ac.uk

**Figure Captions.**

Fig. 1. Interferometry of optomagnetic metamaterials.

(a) The interferometric set-up. (b) Interferometry of a glycerol droplet placed on a glass substrate. The phase of light passing through glycerol lags and the fringes bend to the right. (c) Red light interferometry of a double-dot sample (the edges of the sample are shown by a box). The phase lag is zero and fringes do not bend. (d) Green light interferometry of the structure. The phase of light that passes the medium leads the phase of light passing through the air and the fringes bend to the left. (e) The zoomed version of (d). (f) Green light inerferometry of the structures with electrically connected nanopillars. The lead of the phase in cases (d)-(f) suggests a negative optical path length of the sample.

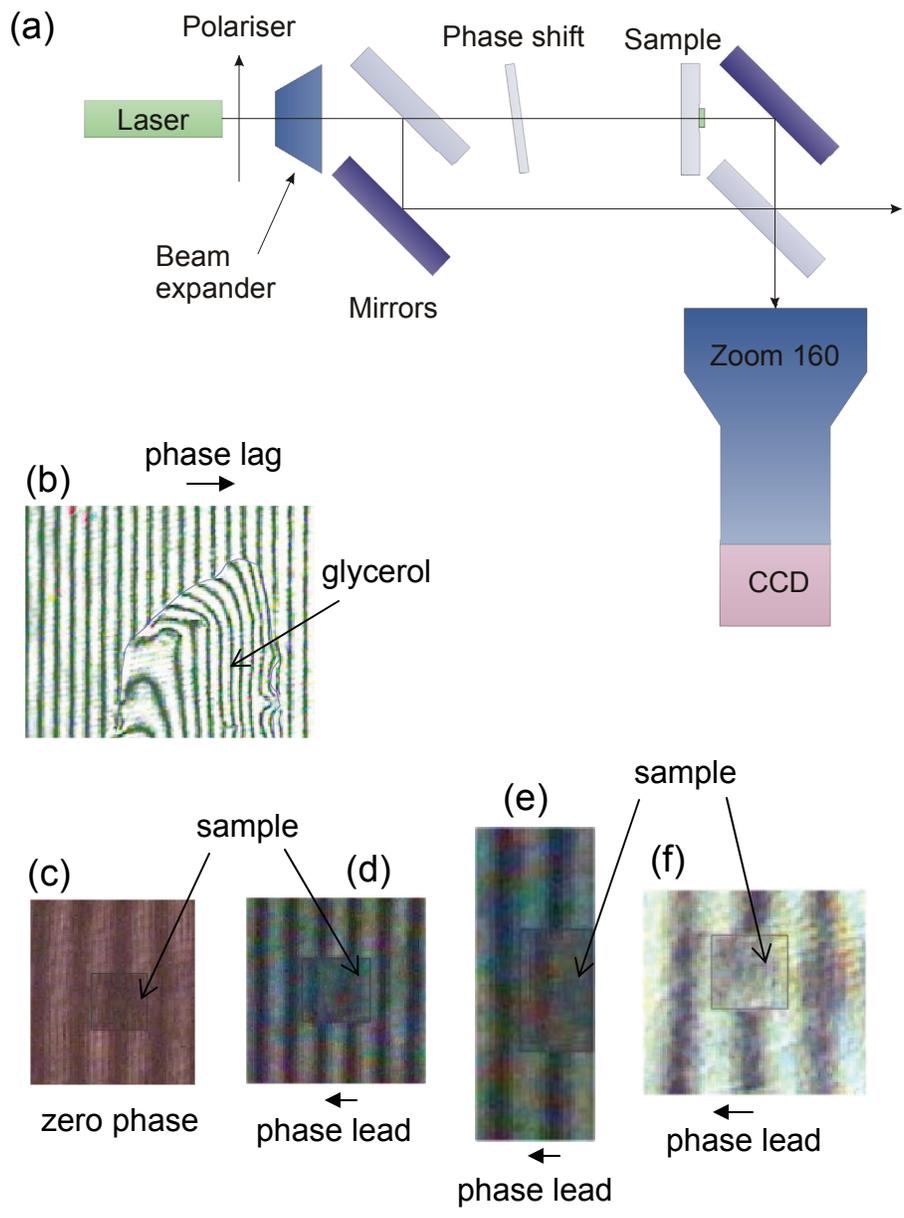

Fig. 1.